# MICOSE4aPS: Industrially Applicable Maturity Metric to Improve Systematic Reuse of Control Software

BIRGIT VOGEL-HEUSER

Institute of Automation and Information Systems, Technical University of Munich, Germany

EVA-MARIA NEUMANN

Institute of Automation and Information Systems, Technical University of Munich, Germany

JULIANE FISCHER

Institute of Automation and Information Systems, Technical University of Munich, Germany

automated Production Systems (aPS) are highly complex, mechatronic systems that usually have to operate reliably for many decades. Standardization and reuse of control software modules is a core prerequisite to achieve the required system quality in increasingly shorter development cycles. However, industrial case studies in aPS show that many aPS companies still struggle with strategically reusing software. This paper proposes a metric-based approach to objectively measure the maturity of industrial IEC 61131-based control software in aPS (MICOSE4aPS) to identify potential weaknesses and quality issues hampering systematic reuse. Module developers in the machine and plant manufacturing industry can directly benefit as the metric calculation is integrated into the software engineering workflow. An in-depth industrial evaluation in a top-ranked machine manufacturing company in food packaging and an expert evaluation with different companies confirmed the benefit of efficiently managing the quality of control software.

CCS CONCEPTS • Software creation and management • Software organization and properties • Real-time systems

Additional Keywords and Phrases: Automated Production Systems, Programmable Logic Controllers, IEC 61131-3, Software Quality

**ACM Reference Format:**

First Author's Name, Initials, and Last Name, Second Author's Name, Initials, and Last Name, and Third Author's Name, Initials, and Last Name. 2018. The Title of the Paper: ACM Conference Proceedings Manuscript Submission Template: This is the subtitle of the paper, this document both explains and embodies the submission format for authors using Word. In Woodstock '18: ACM Symposium on Neural Gaze Detection, June 03–05, 2018, Woodstock, NY. ACM, New York, NY, USA, 10 pages. NOTE: This block will be automatically generated when manuscripts are processed after acceptance

## 1 Motivation and Introduction

automated Production Systems (aPS), i.e., production machines and plants, are highly complex, multi-variant mechatronic systems usually controlled by Programmable Logic Controllers (PLC) programmed according to the standard IEC 61131-3. aPS face increasing challenges, such as high product variability with small lot sizes or a frequently changing product portfolio during an operation time of usually several decades [18, 35, 46] and consequently the need to change the control software accordingly to enable existing mechanics and electrics of such machines to realize different products. The amount of functionality implemented by control software in aPS increases [49]. Typical software tasks comprise, e.g., actuator control for synchronous acceleration and deceleration of drives to move conveyor belts [50]. Since the software is easier to modify on short notice than to exchange automation hardware, changes during an aPS lifecycle are to a large amount implemented via software. Previous industrial case studies [20, 50, 51, 54] show that companies still struggle regarding systematic reuse and modularity of software. The release procedures for control software library modules, e.g., yield a high optimization potential [54]. However, modularity and planned reuse are crucial prerequisites to stay competitive [18] and ensure software evolution and flexibility [7, 14, 36, 40]. There is a high number of tools and research approaches available in the field of software engineering targeting these problems. Nevertheless, there are some well accepted differences between control software in aPS and embedded systems applications [44, 50], including code changes during runtime of the operating machine (feasible with PLCs), hard real-time behavior (implemented by cyclic instead of event-driven software execution on a PLC), and the requirement of less to no knowledge about the operating systems for application programmers as they are primarily experts of the technical process (e.g., for production steps such as forming or pressing) instead of software engineers. Moreover, unexpected environmental conditions or raw material often cause software modifications after the aPS has been built at the customer site, which allows the final integration test only after commissioning. Software product lines were recently researched but also lack applicability for mechatronic systems [20].

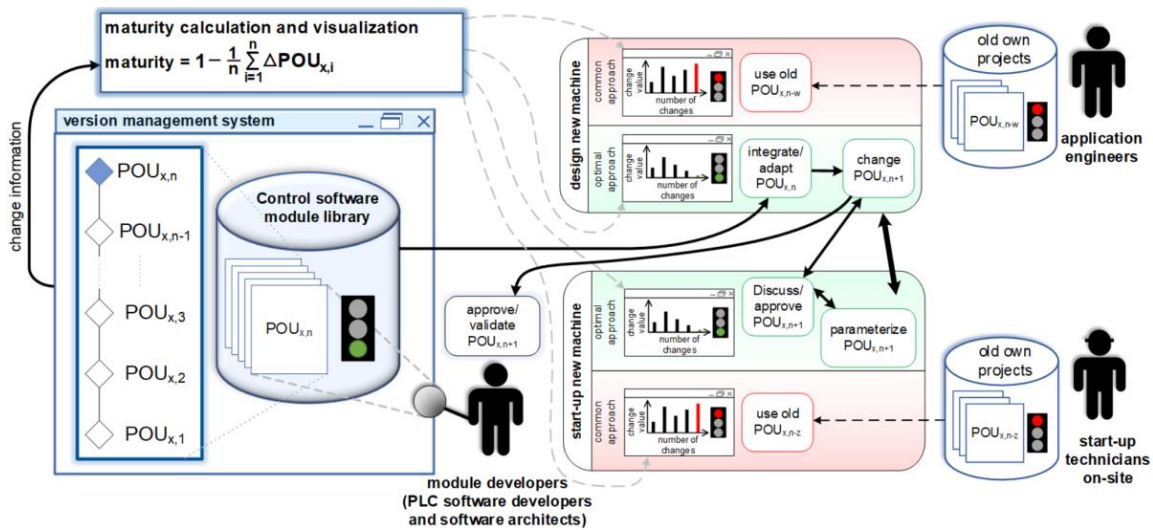

**Figure 1:** Extended workflow for PLC-based control software development based on [52]

A typical workflow for library-based software reuse comprises three phases, i.e., module design by module developers, module use by application engineers, and the start-up on-site by technicians [52] (cf. Figure 1). The module library is designed, maintained, and monitored by module developers, i.e., PLC software developers who program the modules taking care of high quality and reusability by, e.g., standardizing the modules' interfaces and ensuring low coupling to application-specific software parts, and software architects who manage and design the overall software structure [52]. During the design of new machines or their variants, application engineers integrate the library modules into their projects. In practice, however, sometimes code from old projects is used that has already been successfully applied in other applications. In contrast to the library modules, this code was only used once and is less mature. This issue can be addressed by providing developers with an objective maturity indicator highlighting whether a module can be reused without risk. While there are many software metrics in computer science and the embedded systems domain, only a few metrics are available for PLC software. This is due to the significant differences between PLC software and classic embedded systems applications, which refer not only to the pure syntax and data flow of the used programming languages. These differences have been elaborated in previous work [53, 54] and include the often significantly higher overall complexity due to uncontrolled software growth during a lifetime of several decades, the lower qualification level of the commissioning personnel on the construction site, which make object-oriented programming difficult, or significant differences in test capacities. Due to these factors, the quality of PLC software cannot be directly compared to the quality of, e.g., high-level language software from the embedded systems domain, and the interpretability of approaches from computer science applied to PLC software is limited.

Therefore, this paper proposes an approach to objectively measure the <u>m</u>aturity of <u>i</u>ndustrial IEC 61131-based <u>c</u>ontrol <u>s</u>oftwar<u>e</u> in aPS called *MICOSE4aPS*. The approach comprises four parts, i.e., a criticality classification of different software change categories, a maturity metric to objectively measure change impact, a concept to visualize the results for the developer, and an integration to industrial software development workflows to support software reuse in machine and plant manufacturing companies. The industrial benefits of MICOSE4aPS are demonstrated by its integration to the version management tool of Company A, i.e., a world-market leading machine manufacturer for the food and beverage industry, and by discussions with experts from three additional companies. Practitioners can use MICOSE4aPS, e.g., for library maintenance using the metric to identify extensive changes to a software module, thus supporting the software developer in identifying modules that require additional testing effort and in selecting suitable test cases [17]. Besides, MICOSE4aPS can measure a library module's changes during commissioning and thus hint at potential sources of Technical Debt.

The remaining part is structured as follows: First, the current state of the art in analyzing software quality is introduced in Section 2. In Section 3, the requirements to be fulfilled by MICOSE4aPS are derived. Based on the requirements, the approach for maturity calculation is deduced (Section 4). Section 5 describes the evaluation of the approach using a prototypical implementation. Finally, Section 6 provides a conclusion as well as an outlook.

## 2   Background and Related Work on Measuring Code Quality Attributes

In the following, an introduction into the domain of PLC-based control software is provided. The current state of the art regarding metrics for measuring its quality and respective tool support is outlined.

### 2.1 Constraints of the aPS Domain

Usually, aPS are controlled using PLCs [30] characterized by a cyclic program execution due to cyclic deadlines of control tasks to ensure process stability [10]. PLCs are usually programmed according to the IEC 61131-3 defining five programming languages, i.e., two textual ones (Structured Text (ST) and Instruction List), and three graphical ones (Ladder Diagram (LD), Function Block Diagram (FBD), and Sequential Function Chart (SFC)). As a prerequisite for reuse and encapsulation of functionality [50], PLC software is structured into Program Organization Units (POUs), which can be either Programs (PRG), Functions (FC), or Function Blocks (FB). In the following, a *module* represents a POU in the PLC software. The scope and functionality encapsulated within a POU may vary depending on the general software structure, company-specific rules, or applied programming guidelines. Vogel-Heuser et al. [53] derived a five-level module hierarchy for control software in aPS compliant with the ISA-88 physical model [27]. The levels range from *plant modules* controlling whole production plants to *(atomic) basic modules* reading individual sensors or controlling actuators (cf. Figure 2). Depending on the functionality scope, a POU's length may range up to several thousands of source lines of code (SLOC, cf. Table 1).

PLC software development in aPS struggles with challenges that sharply differ from classical software engineering in computer science [50, 54]. Especially the heterogeneity of software structure and quality across companies, including significant differences in the applied programming guidelines and reuse strategies, combined with the fact that established programming paradigms from computer science (e.g., usage of design patterns or object-oriented programming) are not widely used in machine and plant manufacturing [50], lead to the fact that established approaches from computer science are only conditionally applicable to automation software. According to a survey [54], 42% of machine and plant manufacturing companies, e.g., do not yet use the object-oriented extension for the IEC 61131-3.

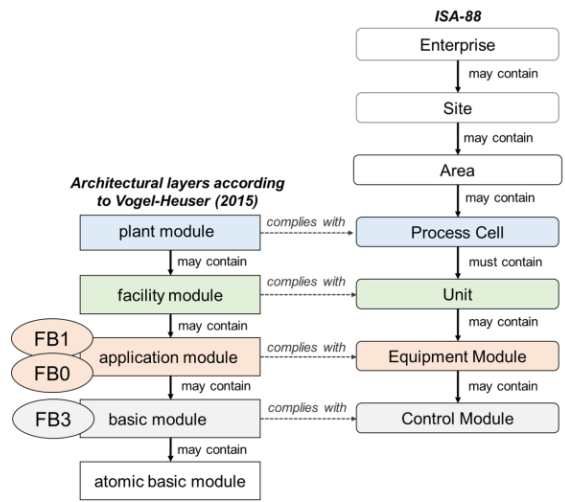

**Figure 2:** Five architectural levels in aPS software according to Vogel-Heuser et al. [16] (left, with FBs from Tab. 1 assigned) mapped to the physical model according to ISA-88 [17]

Version management tools support documenting and restoring the change history of POUs by enabling programmers to generate a desired software state from the *repository*, in which the version history is stored. The desired version from the repository is transferred (*checkout*) to a *working copy*. After the developer has performed a change, the new version is transferred back (*commit*). However, for control software in aPS, version management is not widely spread [20].

**Table 1:** Examples for different FBs from the software of Company A (all FBs implemented in ST, SLOC refers to the last analyzed version)

| Name | Length [SLOC] | Functionality |
|------|---------------|---------------|
| FB0  | 5009          | Integration of a label dispenser by controlling and displaying the dispenser's value |
| FB1  | 1860          | Station for preheating packages to enable their forming |
| FB2  | 2981          | Hierarchy-independent diagnosis FB |
| FB3  | 912           | Reading of a temperature sensor |

## 2.2 Metrics for Measuring Software Quality

Software quality is defined as "the degree to which software possesses a desired combination of attributes" [1] and can be quantified using software quality metrics (in the following referred to as *metrics*), which are defined as "a function whose inputs are software data and whose output is a single numerical value that can be interpreted as the degree to which software possesses a given attribute that affects its quality" [1]. Metrics make software measurable and thus manageable [15]. In the field of embedded systems, it has been identified that "the embedded software design cycle can greatly take advantage from the use of software product quality metrics" [41].

Using metrics, quality assessment becomes less subjective. However, metrics do not aim at replacing but at supporting the human to identify potential weaknesses in the software. In the following, the quality attribute *maturity* is used as defined by ISO 25010 [28], i.e., the "degree to which a system, product or component meets needs for reliability under normal operation." For aPS, the regular operation also includes enabling subroutines such as the manual operation of selected drive axes to restart a machine, e.g., after an error has been resolved. Maturity is, therefore, essential for aPS software, which has to fulfill safety-critical functions. However, a mapping study with more than 150 papers on assessing code quality attributes [5] revealed that maturity is one of the least studied quality attributes.

In computer science, a variety of metrics is available [16]. Lines of Code (LOC) [32] measures software size by counting the code file lines. Often only non-commentary, non-empty lines are of interest (SLOC [45]). As the IEC 61131-3 also comprises graphical languages, the transferability is limited. Halstead [9, 24, 32] assesses code complexity based on the number of operands (passive elements, i.e., values) and operators (active elements manipulating values). McCabe [32, 37] calculates a program's complexity based on the independent paths through the control flow graph. Henry and Kafura [25] introduce the metrics Fan-In and Fan-Out determining a software unit's incoming and outgoing information flows.

For PLC software, metrics for quality assessment are less common than in computer science [43]. Some approaches define metrics for IEC 61131-3-languages (e.g., LD [33] or ST [34]) for measuring program length, difficulty, or cognitive complexity. Further metrics measure the reusability and reliability of POUs [39] but consider POUs as encapsulated units and do not consider their implementation. However, a detailed implementation analysis of POUs is crucial to analyze module libraries.

There are approaches for maturity analysis of open-source software [26] or reusability assessment of Java components by investigating their coupling [23]. These approaches cannot be directly applied due to the high amount of input and output parameters of PLC software in aPS.

The assessment of software changes gives valuable insights regarding software evolution [21] and may indicate implicit architectural dependencies [48] or defect-prone software components [38]. Using machine learning approaches, change analysis can even be used for measuring the effort of code changes [58]. However, additional adaptions would be required to analyze changes in PLC software in the context of assessing its maturity.

Although existing metrics such as Halstead's program length and McCabe's Cyclomatic Complexity can be adapted to IEC 61131-compliant control software [13, 57], it is generally challenging to transfer metrics from computer science that go beyond calculating scope and complexity and determine specific quality attributes, such as maturity. This is because, in addition to a frequent focus on object orientation and a lack of options for distinguishing between different change criticalities, these metrics usually cannot be parameterized for different company-specific software characteristics. These company-specific differences are illustrated using the result of an industrial questionnaire study from previous work [51]: In the study, two companies, in particular, stand out by achieving top scores in different categories (modularity, testing, and operation). However, the software projects' detailed analysis clearly shows distinctive structural differences (cf. Figure 3). This illustrates the heterogeneity of control software in machine and plant engineering and shows that software quality cannot be determined with a generic formula based on syntax and data flow but that adjustable parameters are needed to consider company-specific strategies.

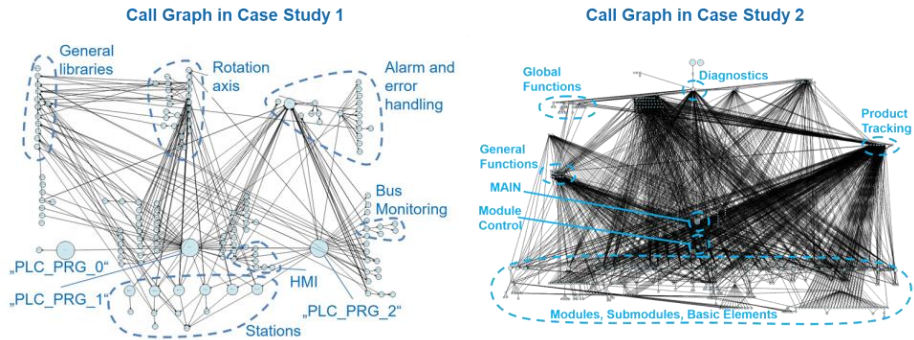

**Figure 3: Comparison of call graphs and functionality distribution in two machine manufacturing companies [51]**

Vogel-Heuser et al. [52] propose an equation to calculate the maturity of FBs based on the number of different change types *n* and the respective change scope ($\Delta FB_x$, equation (1.1)) made to an individual $FB_x$, serving as the groundwork for this paper:

$$Maturity = 1 - \frac{1}{n} \cdot \sum_{i=1}^{n} \Delta FB_{x,i} \quad (1)$$

With:

$$\Delta FB_{x,i} = \frac{\sum changed\ items\ of\ type\ i}{\max\{\sum items\ of\ type\ i\ before; \sum changed\ items\ of\ type\ i\}} \quad (1.1)$$

It is distinguished between eight possible change types, including change of hardware interfaces or module implementation, which is determined by counting the changed SLOC before and after a change. A traffic-light-based approach to visualize the maturity values is introduced (cf. Figure 1). Two of the four proposed maturity indicators [52] will be investigated in detail in this paper:

- **FBs at lower call levels are less affected by changes than FBs at higher levels (cf. Figure 2):**
FBs from lower levels can be standardized more easily because they usually fulfill tasks such as controlling a simple drive, making them less application-dependent. Therefore, they are reused more frequently, increasing the probability of detecting errors, and the need for adaptions is usually low [51]. Additionally, comprehensive changes are usually avoided due to the high amount of reuse, leading to a high risk of cross effects to the calling FBs (cf. equation (2)). The formulas in the following refer to the change scope ($\Delta FB$) and represent, therefore, the (simplified) reverse value of the maturity (cf. equations (1)).

$$\Delta FB_{x,level} > \Delta FB_{x,level-1} > \Delta FB_{x,level-2} \quad (2)$$

- **The amount of changes to an FB decreases as it proceeds in its lifecycle:**
The release procedure of library modules is usually an iterative process accompanied by testing routines to verify the changes [51]. Most of the library module changes usually occur during an FB's initial design process, but also in later phases when the customer is already operating the machine, software changes are necessary (e.g., small adaptions due to an exchange of hardware parts). However, as changes are more challenging to handle in later phases, the change impact should be kept at a minimum (cf. equation (3)).

$$\Delta FB_{x,operation} < \Delta FB_{x,Start-up} < \Delta FB_{x,Design} \quad (3)$$

The approach is a first attempt to determine the maturity of library modules in aPS based on the scope of change. However, it has been evaluated only manually for a lab-sized demonstrator (less than 100 SLOC). To enable applicability in an industrial environment, the equation needs to scale up, and the calculation procedure needs to be automated and integrable in an industrial software engineering workflow with its tools.

## 2.3 Tool Support for Measuring Code Quality

Appropriate tool support is a mandatory prerequisite to measure code quality [28] in an industrial context. Although classical code development tools already exist for static code analysis [6, 12, 31, 56], IEC 61131-3 is so far only supported by a few suppliers [3], e.g., [2, 8, 11, 29]. Prähofer et al. [42, 43] developed an analysis tool for textual IEC 61131-3 languages, including naming conventions, complexity, and metrics such as LOC, but do not consider the evolution of POUs based on underwent changes or the application of maturity metrics.

Different frameworks are available, e.g., for evaluating IEC 61131-3-compliant control software using Semantic Web Technologies [19] or for SFC code [47], but these frameworks focus on the static structure of the control software and do not provide support to analyze software evolution and change criticalities.

There are approaches for using data from version management tools for assessing software quality, e.g., by combining versioning and bug tracking data [21] or by identifying architectural weaknesses by investigating the change history of software classes [22]. However, these approaches do not provide a detailed change analysis on the code level. Therefore, the granularity of the approaches [21, 22, 58] is not sufficient for a detailed maturity analysis of industrial IEC 61131-3-compliant control software.

## 3 Requirements for MICOSE4aPS

To derive the requirements of MICOSE4aPS, a preliminary study is performed to evaluate the performance of equation (1.1) in an industrial context. Therefore, the initial approach [52] is applied to the change history of an exemplary $FB_0$ of Company A (cf. Table 1), exported from the applied version management tool. $FB_0$ was selected together with PLC experts of Company A. It has been frequently affected by various changes within the last two years and, therefore, provides a representative range of different change types and scopes.

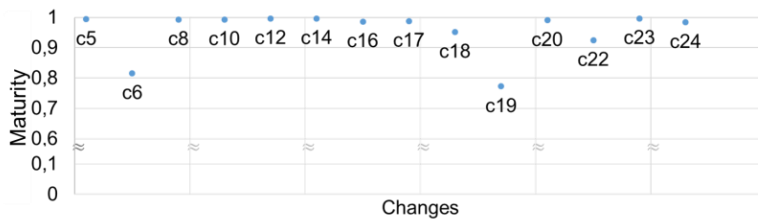

Figure 4: Resulting maturity development of $FB_0$ by applying equation (1.1)

Discussing the results with PLC experts of Company A reveals three challenges:

*Challenge 1*: Deriving an implementation change's criticality based on the number of changed SLOC is not sufficient since a change of a line containing a simple assignment, e.g., is less error-prone compared to a line containing parts of a loop.

*Challenge 2*: In the case of large POUs such as $FB_0$ (several thousand SLOC), the implementation changes almost do not affect the maturity, even in case significant parts of the code were refactored (e.g., $c_6$ in Figure 4: comprehensive refactoring, but maturity > 80%).

*Challenge 3*: The calculation procedure is not beneficial in an industrial context: The values were not calculated automatically, and the proposed visualization, i.e., a traffic light system illustrating a single calculated maturity value [52], is not sufficient to rate an FB's reusability as these values only represent snapshots of a specific change's impact on the maturity.

From these challenges, requirements for enlarging the approach for maturity assessment of control software in aPS, i.e., MICOSE4aPS, are derived in the following.

### *R1. MICOSE4aPS should deliver reliable results for industrial control code.*

As metric (1.1) did not lead to reliable results when applying it to industrial PLC code (cf. Challenge 1), the calculation must be enlarged. Therefore, two directions need to be considered covered by R1.1 and R1.2:

*R1.1. MICOSE4aPS reliably leads to high maturity values in case of changes with low criticality.*

*R1.2. MICOSE4aPS reliably leads to low maturity values in case of changes with high criticality.*

*R1.3. MICOSE4aPS should be able to reflect different factors that influence change impact.*

Requirement R1.3 refers to the fact that software maturity is affected by the way of committing changes. MICOSE4aPS should reflect the following influencing factors on the change impact:

- *Applied Change Strategy*: Discussion with industrial PLC programmers revealed that POUs are treated differently depending, e.g., on their size or reliability, which also affects the way of implementing changes.
- *Change Category:* It is assumed that if a new feature enlarges a POU's functionality, this is accompanied by a significant change to the code leading to an increased risk of introducing faults, which hamper the POU's ability to meet needs for reliability under normal operation. Therefore, the implementation of additional functionality should lead to higher change impacts compared to, e.g., a bug fix (cf. equation (4)).

$$\Delta POU_{x,BugFix} < \Delta POU_{x,NewFunctionality} \quad (4)$$

- *Number of changes made to a POU before*: Based on the assumption that changes are made to enhance the software by fixing errors or enlarge its functionality, it is derived that a POU is improved on average the more often it has been changed without adding functionality (cf. equation (5) with c = the total number of changes made to a POU during its change history).

$$\Delta POU_{x,c-1} < \Delta POU_{x,c} < \Delta POU_{x,c+1} \quad (5)$$

- *Lifecycle phase:* The amount of changes is expected to decrease as a POU respectively the aPS controlled using the POU proceeds in its lifecycle (cf. equation (3)). As the central part of a POU's functionality is expected to be specified during its design phase, it is expectable that the significant share of *functional* changes occurs during this phase and decreases in later phases.

$$\left(\frac{\Delta POU_{x,funct}}{\Delta POU_{x,total}}\right)_{Design} > \cdots > \left(\frac{\Delta POU_{x,funct}}{\Delta POU_{x,total}}\right)_{Operation} \quad (6)$$

- *Architectural level:* As stated by indicator (2), POUs are affected by different change impacts depending on their architectural level (cf. Figure 2), e.g., due to different standardization degrees. The metric is required to distinguish these differences.

### *R2. MICOSE4aPS should be adaptable for different boundary conditions.*

R2 addresses the lack of scalability of equation (1.1) to industrial PLC code (cf. Challenge 2). To validate and specify R2, the following sub-requirements are introduced:

*R2.1. MICOSE4aPS should be scalable to different sizes of POUs and scopes of change.*

The initial equation (1.1) is not precise enough to distinguish different change types and their impact on different types and sizes of POUs (cf. Table 1). MICOSE4aPS is expected to solve these issues.

*R2.2. MICOSE4aPS should be adaptable to different company-specific constraints.*

Depending on the company, the control software is affected by various influencing factors, such as different PLC platforms or IEC 61131-3 programming languages. To enable usability on a larger scale, MICOSE4aPS should be applicable across companies without significant adjustment effort.

*R3. The application of MICOSE4aPS in an industrial context should beneficially support the module developer.*

The initial approach [52] is not beneficial in an industrial environment due to missing tool support and insufficient visualization (cf. Challenge 3). Therefore, MICOSE4aPS should fulfill the following sub-requirements:

*R3.1. The results of the maturity metric and their visualization should be understandable for module developers.*

To create actual benefit for PLC software developers and software architects, the resulting maturity values are required to be intuitively understandable.

*R3.2. MICOSE4aPS can be integrated into a company's software development workflow.*

To use the maturity assessment in an industrial environment, it should be simple to integrate MICOSE4aPS into the version management tools and workflows of different companies.

# 4 Approach for Maturity Calculation of Control Software Library Modules

This section derives the enlarged approach MICOSE4aPS for maturity calculation of control software modules.

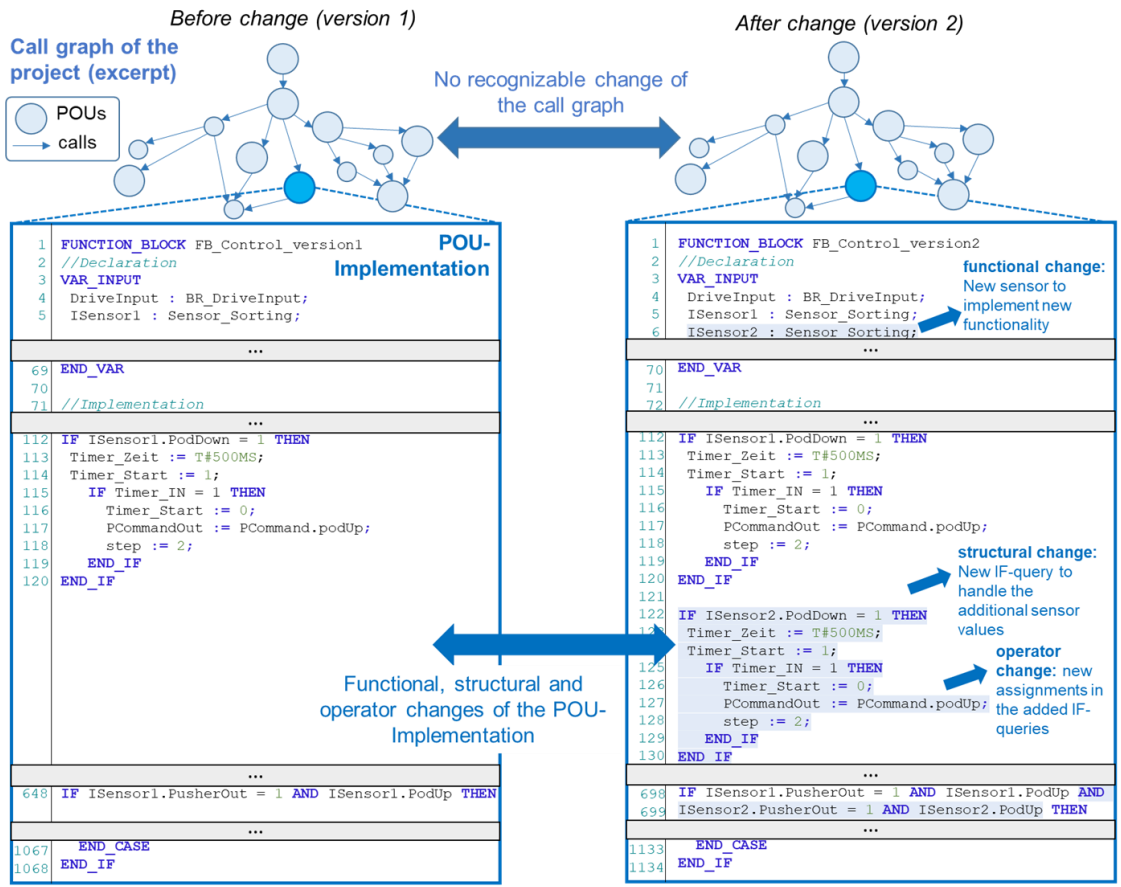

**Figure 5:** Excerpt of code changes on the FB "FB_Control": Enlargement of functionality and of an FB controlling a sorting system (left: before change, right: after change)

## 4.1 Categories of change criticality

In the following, a code example for an evolving POU is used to demonstrate why established metrics from computer science are not sufficient for precise change estimation on a software unit with a typical scope and complexity in the field of machine and plant engineering. The "FB_Control" controls the sorting system of a demonstrator plant for transporting, sorting, and filling bottles with granulate comprising more than 1,000 SLOC. The following example describes a typical change scenario in aPS: To optimize the sorting process of the bottles, an additional sensor is attached to one of the conveyors (ISensor2), i.e., an additional hardware input is included to implement the enlarged system functionality *(functional change)*. This change entails further changes of the implementation's structure *(structural changes)*, i.e., changes affecting the control flow branches of the POU, e.g., additional loops or conditions to adapt the behavior to the new sensor input. Besides, the change leads to further fine-granular changes at the operator level, e.g., assigning timer values *(operator changes)* that do not affect the control flow branches of the implementation, which makes them less critical compared to functional and structural changes. The POU's implementation changes are conducted using *Copy, Paste & Modify (CPM)*, which is accompanied by the risk of missing necessary adaptions and, thus, introducing faults to the code [20]. From a developer's perspective, it would be helpful to capture this risk using a suitable metric. Applying established metrics to this scenario (cf. Table 2) reveals that none of them shows a significant increase: The largest increase of 9.2% shows the metric Cyclomatic Complexity, which, however, is still below the threshold of 10% for non-critical changes from [52], which is considered as an intuitive threshold by industrial experts. This result is not surprising since Halstead's and McCabe's metrics focus on complexity, which is only marginally changed in this example. Metrics evaluating the incoming and outgoing information flow of a POU, i.e., Fan-In and Fan-Out [25], stay constant because the change only affects the POU's implementation, not its use by its environment, which is also apparent from the call graph staying unchanged (cf. Figure 5, top).

**Table 6: Results of established metrics for the change scenario (cf. Figure 5)**

|  | FB_Control_version1 | FB_Control_version2 | Difference |
|---|---|---|---|
| **Lines of Code [45]** | 1068 | 1134 | ↑ 6.2% |
| **McCabe Cyclomatic Complexity [37]** | 239 | 261 | ↑ 9.2% |
| **Halstead Difficulty [24]** | 175,78 | 188,22 | ↑ 7.1% |
| **Fan-In [25]** | 4 | 4 | → 0% |
| **Fan-Out [25]** | 189 | 189 | → 0% |

Summarizing, established metrics from computer science are not capable of representing critical changes on large-scale POUs in aPS as they do not distinguish differences in change criticality and cannot be customized to the POU scope or company-specific criteria for change criticality. To consider the impact of different change types in the metric by appropriate weighting factors, the criticality categories mentioned above are used, i.e., functional (highest criticality), structural (medium criticality), and operator changes (low criticality). In total, 69 change terms distributed to three different criticality categories are distinguished. Since ST has the broadest language scope and is preferred by most PLC developers, the structural and operator changes focus on ST. However, the evaluation shows the applicability also for graphical languages.

## 4.2 Derivation of the calculation rules

In the following, an optimized calculation rule for the change terms $\Delta FB_x$ in metric (1) is derived to solve the scaling problem (cf. Challenge 2). A combination of an exponential and a linear term is empirically proposed. Previous evaluation results [52] show that a linear change term calculation leads to good results for small POUs but is not applicable for larger POUs (cf. Section 3). Hence, an additional exponential summand is introduced (line 2 in (1.1+)).

$$\Delta POU_x = k_l \cdot w \cdot \frac{\sum changes}{\max\{\sum before; \sum changes\}} + k_e \cdot w \cdot \left(1 - \exp\left(-p \cdot \left(\frac{\sum changes}{\max\{\sum before; \sum changes\}}\right)\right)\right) (1.1+)$$

Metric (1.1+) applies to all three POU types but mainly addresses FBs and FCs, which can be standardized in POU libraries, whereas PRGs are mainly used as entry point of the software project to trigger the execution. Applying only the exponential term of (1.1+) ($k_l$=0 and $k_e$=1), smaller modules are accompanied by a higher probability of ending up in the range of higher change ratios (cf. Fig. 6) as for smaller FBs, the absolute number of operators is lower. Therefore, depending on the POU size, the two summands of (1.1+) must have a different share, which is set by the parameters $k_e$ (greater for *large* POUs) and $k_l$ (greater for *small* POUs). In total, $k_e$ and $k_l$ must sum up to 1:

$$k_l + k_e = 1 \text{ (6)}$$

Based on previous experience gathered by evaluating PLC code of different companies from the aPS domain [51, 55], the following thresholds are proposed: POUs smaller (larger) than 150 (1,000) SLOC are considered as small (large). Medium size modules comprise between 151 and 999 SLOC, and for this range, the ratio of $k_l$ and $k_e$ is received using linear interpolation leading to the following result:

$$k_e = \begin{cases} 1, \text{ if } \#SLOC \geq 10^3, \\ \frac{1}{850} \cdot (\#SLOC - 150), \text{ if } 150 < \#SLOC < 10^3, \\ 0, \text{ if } \#SLOC \leq 150 \end{cases} \quad (7)$$

$$k_l = 1 - k_e, \text{ respectively.}$$

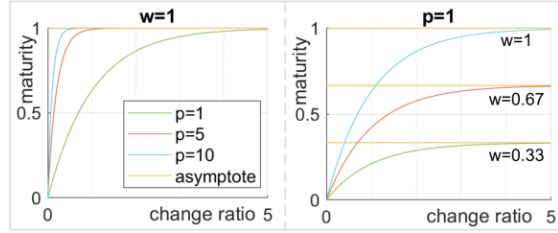

**Figure 6:** Illustration of the curve progression of equation (1.1+) with different variations of p (left) and w (right) in case of large modules (k$_l$=0, k$_e$=1)

The exponential part of (1.1+) solves the scaling problem for larger modules as it more sensitive regarding smaller changes (cf. Figure 6): For change ratios near zero, the curve rises sharply, whereas for higher change ratios, the curve flattens. The parameter $p$ specifies how fast an increasing change ratio leads to an asymptotic approximation (cf. Figure 6, left). It is assumed that the maturity approximates zero in case of change ratios near one (approximated by 0.99, as 1 cannot be reached), which is fulfilled by $p$ = 5:

$$\Delta POU_x = 1 \cdot (1 - \exp(-5 \cdot (1))) \approx 0.99 \quad (8)$$

The weighting factor $w$ represents the change criticality, i.e., it specifies the maximum change value that a specific change term can reach (cf. Figure 6, right). The factor $w$ is composed of two differently weighted summands, i.e., a level summand $s_1$ representing the impact of the change according to its change category and a summand $s_2$, which allows companies to adjust the weighting of individual change terms according to, e.g., the applied programming rules. As $s_1$ is supposed to specify the main part of a change term's criticality and $s_2$ represents company-specific refinement, both summands are weighted according to the Pareto Distribution [4]:

$$w = 0.80 \cdot s_1 + 0.20 \cdot s_2 \quad (9)$$

Regarding $s_1$, an empirical approach is to evenly distribute the weighing factors of the three levels leading to the following results:

- Functional changes: $s_{1,FC}$ = 1
- Structural changes: $s_{1,SC}$ = 0.67
- Operator changes: $s_{1,OC}$ = 0.33

The company-specific value $s_2$ needs to be specified for each change term, e.g., by PLC programmers rating the changes according to their experience.

Figure 7 summarizes the steps of applying the metric based on the change scenario introduced in Figure 5. Overall, the change leads to a maturity of about 81%, i.e., a change criticality of almost 20%, which corresponds much better to the criticality of the *CPM* changes caused by the new hardware variable compared to the results of the analyzed complexity metrics from computer science (cf. Table 2). The accuracy of the values is demonstrated in Section 5 as part of the industrial evaluation.

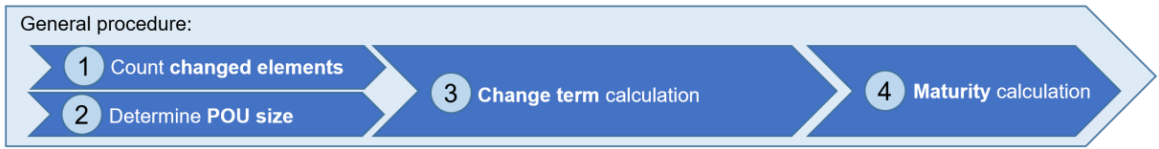
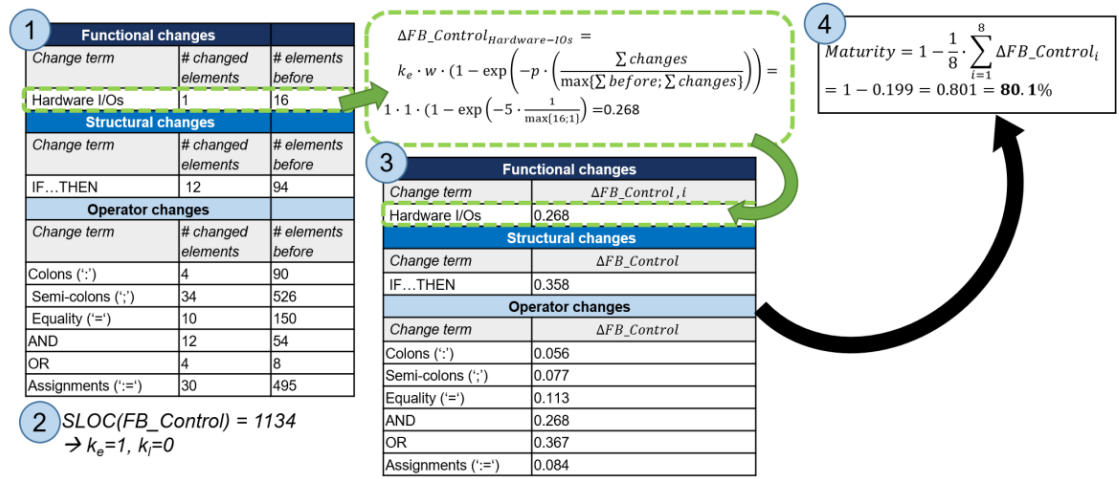

Figure 7: Procedure of metric application and demonstration for change scenario of FB_Control (cf. Figure 5)

## 4.3 Maturity Visualization

To provide a framework for assessing the calculated maturity value, an appropriate visualization is necessary (cf. R3.1). Additional information on preceding changes and their impact on maturity should be displayed. The implementation of the resulting visualization concept is depicted in Figure 9. The threshold values for the traffic light system are adopted from previous work [52].

# 5 Evaluation of MICOSE4aPS

In this section, the use case to evaluate the validity, applicability, and benefit of MICOSE4aPS in an industrial context (Section 5.1) and the evaluation results (Section 5.2) are derived and discussed (Section 5.3).

## 5.1 Industrial implementation in Company A

Company A is a global leader in packaging machines and was chosen as it achieved the best results in a preceding survey on software modularity, quality assurance, and operation, and the software architecture did already undergo analyses [51]. Therefore, appropriate module quality for the validation of MICOSE4aPS can be assured. Additionally, the results of MICOSE4aPS can be compared to Company A's previous evaluation results to increase validity.

The control software is mainly written in ST and is structured according to ISA-88 [27]. The highest level is represented by a single FB calling FBs from the next lower level, i.e., the Stations controlling individual processing steps. Underneath, substations fulfill hardware-oriented control tasks, e.g., controlling a label dispenser (cf. Figure 8, Table 1).

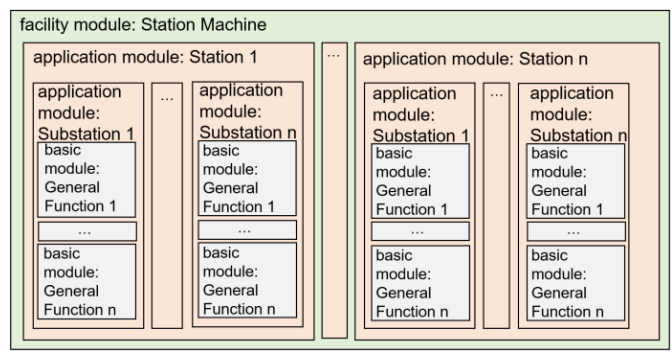

Figure 8: Module hierarchy of Company A according to levels of Vogel-Heuser et al. [53]

To validate R3.2, i.e., the integration of MICOSE4aPS to an industrial software development workflow, the derived approach (cf. Section 4) is linked to the version management toolchain of Company A. The company uses Mercurial, which is an open-source system written in Python. A Python script is developed to access the change data from Mercurial to calculate maturity values. The metric results and additional information about the changes extracted from Mercurial are stored as JSON files into a document-oriented database (MongoDB).

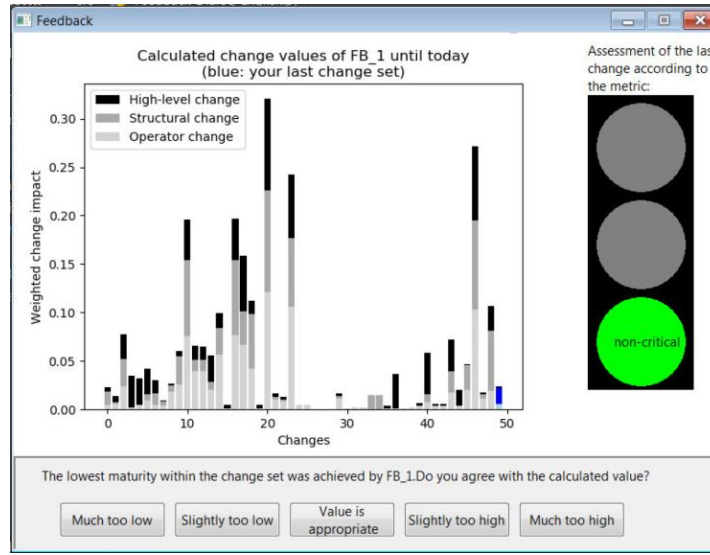

**Figure 9:** Implemented user interface to visualize the calculated maturity and previous change impacts

In case of a change request, the programmer who initiated the request sets the change category according to their experience. There are no strict rules on assigning which category. However, there are coarse guidelines to select an appropriate change category: *Enhancements* represent the least critical category and usually comprise small improvements of the code. *Bug Fixes* aim for fixing a single error and represent a more critical interference with the code than Enhancements, but the proportion of changed SLOC should be kept at a minimum. *New Features* represent new functionalities, which makes them significantly more critical, and they usually affect larger proportions of code. *New Developments* are the most extensive change category and comprise several Sub-Features, which multiple programmers usually develop during a longer period. In case of a change request, all changed files are committed by the programmer as a *changeset*. The change category refers to all files in a changeset.

To customize the metric to Company A, the weighting factors $s_2$ were determined using a points score to rate each change term's criticality within a workshop with three PLC programmers.

A user interface (cf. Figure 9) was developed to visualize the resulting maturity values using a graphical user interface (GUI) toolkit of Python (wxPyton). In addition to the change history and the traffic light indication, the interface provides further information by mouse-over events, e.g., a legend for the different change criticalities. The Python implementation was integrated to Mercurial and is called after a changeset commit is finished. The developer receives direct feedback on the criticality of the change. The calculation is implemented concurrently to ensure that the developer can continue his work without time restrictions. The concept derived in Section 4 combined with the implementation form the four-step approach MICOSE4aPS (cf. Figure 10).

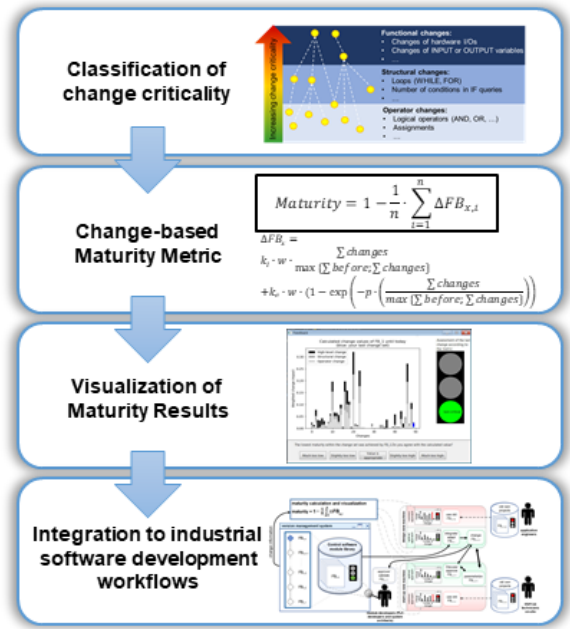

**Figure 10:** Resulting approach MICOSE4aPS

## 5.2 Validation of MICOSE4aPS

To validate whether MICOSE4aPS meets the requirements formulated in Section 3, the calculated maturity of individual FBs was evaluated in joint discussions with software architects from industry and academia. Additionally, the calculation results were displayed to PLC software developers after committing a change using the feedback interface (cf. Figure 9). They were asked whether the values comply with their expectation. Furthermore, it was investigated whether MICOSE4aPS can be applied in other companies.

**Validation of the metric's reliability (R1)**

First, MICOSE4aPS was evaluated independently by four experts from industry and two from academia. The calculated maturity of nine FBs of different sizes and architectural layers was discussed by comparing the calculated changes using the metric (1.1+) with the actual changes made to the code. In the following, the applied procedure is demonstrated using $FB_1$ (cf. Table 1, Figure 11). The $c_i$ in Figures 11, 12, and 14 refer to the change number in the respective FB change history extracted from the version management system. The numbers are not continuous because changes, which are not identified by MICOSE4aPS (e.g., a change of comments), were not included.

Besides one outlier (c27), the maturity of $FB_1$ is always higher than 80%. For c27, several functional changes were made to $FB_1$, e.g., one internal and two interface variables were added. Additionally, large amounts of the FB's structure were changed by adding (partly nested) IF-loops. Several FB calls were added using *CPM* of existing calls. Hence, both industry and academia experts concluded that the calculated maturity (62%) is appropriate.

Changes c11 and c38 achieve medium maturity values (c11: 83%, c38: 85%). For c11, this is due to several new internal variables, and some added IF-loops, partly also containing new FB calls. The lower maturity of c38 is mainly caused by adaptions of existing IF-loops affecting the number of conditions and the nesting depth. The experts concluded that both maturity values are appropriate.

The remaining maturity values were checked content-wise against the performed changes on a random basis leading to the result that the high maturity is appropriate.

Regarding the comparison of the absolute number of changes (Figure 11, top) with the weighted change impact (Figure 11, bottom), changes c33 and c34 stand out as they are barely visible in the absolute view but show relatively high change impacts in the weighted view. According to the experts, this is appropriate as in both cases, a new interface variable was added (functional change) and (1.1+) scales correctly.

Next, the feedback interface (cf. Figure 9) was activated for seven PLC software developers, and the dialogue was displayed each time they committed a change. In addition to the values gathered using the interface, the developers were asked in person for their assessment reasons. The developers confirmed the results of MICOSE4aPS, and, therefore, the validation with both types of experts (software architects and PLC software developers) confirmed the reliability of MICOSE4aPS: Metric (1.1+) reliably detects comprehensive changes through low maturity values (R1.2) whereas non-critical changes lead to high maturity values (R1.1).

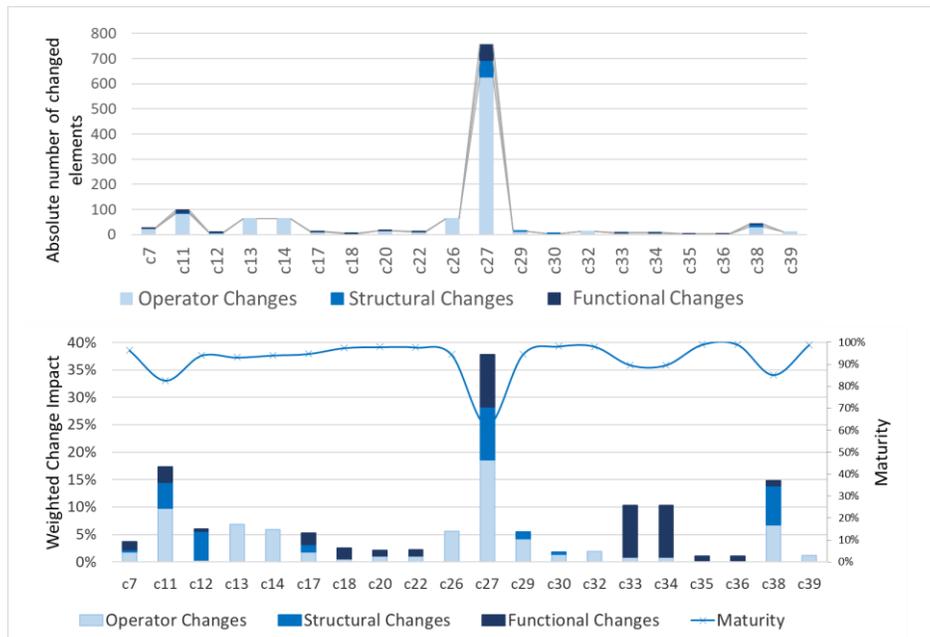

**Figure 11:** Comparison of the absolute number of changed operators per level (top) with calculated changes on the three levels using metric (1.1+) (bottom) for $FB_1$

To investigate whether MICOSE4aPS meets requirement R1.3, i.e., that different influencing factors on change impact are reflected, it is analyzed how a module's size and reliability affect its change behavior:

*Change strategy depending on module size:* Software developers of Company A observed a correlation between large module size and high internal complexity, e.g., due to comprehensive nested IF statements. Therefore, even small changes can lead to potential cross-effects on other parts of the POU, which may harm the existing functionality. On the other hand, small modules often show a lower internal complexity as the small size does not allow large, complex code structures. To analyze potential effects on the software maturity, changes of four FBs with short program length (100 – 200 SLOC) were evaluated with the software architects. Changes to the smaller FBs were very comprehensive, ranging from extensive structural changes to new hardware interfaces, reflected by lower maturity values (1.1+).

*Change strategy depending on module reliability:* The software architects stated that if an FB reliably fulfills its functionality over a long period, it is recommended to avoid making critical changes to it, e.g., by implementing a new feature. Discussions with the responsible software architect of $FB_2$ (cf. Table 1), e.g., revealed that $FB_2$ is frequently instantiated since 2008, proving a reliable functionality, which should not be compromised by changing it. This specification is reflected by metric (1.1+) by consistently high maturity values (cf. Figure 12).

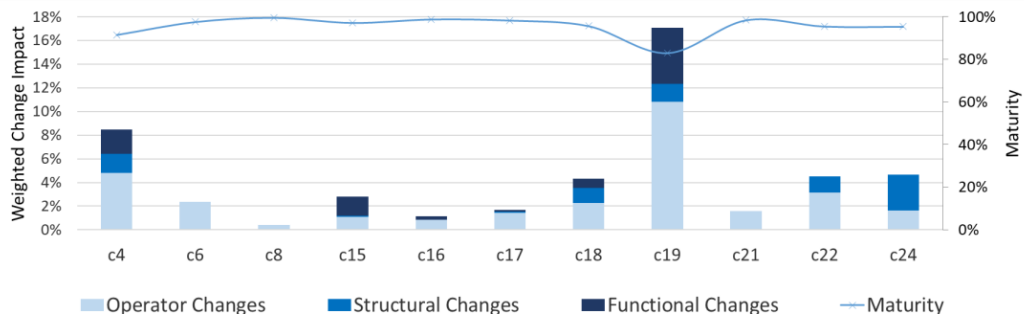

**Figure 12:** Weighted change composition according to metric (1.1+) for $FB_2$

*Change Category:* To investigate the influence of different change categories on the maturity value (cf. equation (4)), the average maturity and the standard deviation for the four change categories in Company A are determined (cf. Table 3). In total, 856 changesets were evaluated with an average maturity value of 94%. This complies with previous survey results [51] confirming the high maturity of Company A's software. Moreover, only changes starting from 2015 are analyzed, which means that the design phase is not captured for all FBs, in which the most extensive changes are expected to be made. New Developments are usually programmed by committing multiple small changes over a long time, whereas New Features are often committed as one significant change. Therefore, higher maturity values for New Developments are reasonable. Enhancements and Bug Fixes show the highest maturity values (94% both) due to the low change criticality affecting only a few LOC. Although the differences are small due to the company's high overall maturity, they can be identified using metric (1.1+). MICOSE4aPS can be used to perform consistency checks, in case, e.g., a bug fix leads to remarkable low maturity values.

**Table 3:** Maturity evaluation with standard deviation σ of two years classified into the four change types (changeset numbers in the individual categories do not sum up to 856 as some changes do not belong to one of the types (e.g., adding of new tags))

|  | Maturity | σ | #Change Sets |
|---|---|---|---|
| *All changesets* | *94%* | *0.082* | *856* |
| **Enhancements** | 94% | 0.079 | 105 |
| **Bug Fixes** | 94% | 0.062 | 462 |
| **Features** | 91% | 0.10 | 215 |
| **Developments** | 93% | 0.098 | 42 |

*Number of changes made to a POU before:* Sorting the maturity of all changes during the last two years by the change number counted from 2015 (cf. Figure 13) reveals that the scatter of the maturity values decreases with an increasing change number. The maturity range decreases and settles to relatively high results. Hence, MICOSE4aPS reflects the influence of a rising change number.

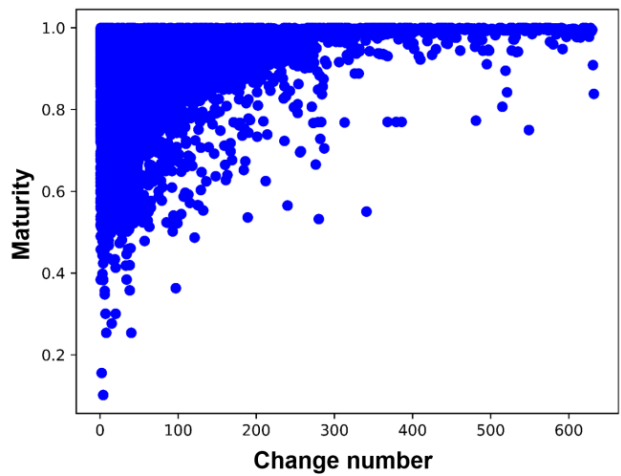

**Figure 13:** Maturity development of all changesets during the last two years plotted against the number of the performed change

*Lifecycle phase:* To investigate whether the amount of functional changes decreases as the POU proceeds in its lifecycle (cf. equation (3+)), the change development of an exemplary $FB_3$ (cf. Table 1) starting from the initial development of the module until its operation is investigated (cf. Figure 14).

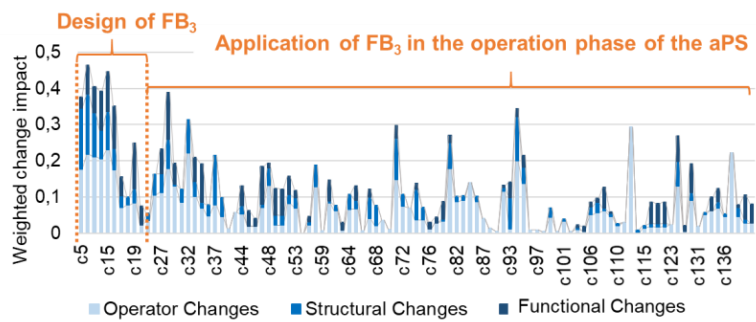

**Figure 14:** Maturity development of an exemplary FB during its whole lifecycle (February 2004 until June 2018)

During the initial development phase, the maturity varies widely, and several changes are leading to low maturity values (e.g., c5-c16, c28). In later phases of the considered period, some significant outliers can be identified (e.g., c94, c113, c124), but the frequency of low maturity decreases. In the early phases of the FB's lifecycle, the proportion of both functional and structural changes is remarkably high. In later phases, operator changes are dominant, and functional changes cause only a small impact. It is reasonable that there are many functional and structural changes in the early phases, as the main parts of the POU's functionality are developed and adapted. In later phases, the basic functionalities of the POU are nailed down, and it is only concerned by changes mainly on the operator level. As expected, the total change impact decreases with the number of performed changes. A senior engineer of Company A described the displayed maturity development as typical: After the series launch (around change 22), the maturity first decreases, as many faults are only discovered after the operation phase has started. Afterwards, the maturity course shows a wavelike behavior due to the implementation of New Features followed by "Bug Waves" caused by introducing new functionality. MICOSE4aPS reflects the influence of the POU's lifecycle phase within the context of the considered company. However, only one FB was evaluated, which limits the validity of the observation.

*Architectural level:* Sorting the maturity values by their architectural level reveals that the maturity results on different change levels are very similar. Only the General Functions show a slightly lower average maturity of 93%. Analyzing the FBs on the General Function level shows that these are smaller (partly only 100 to 200 SLOC) than FBs from higher levels, as they usually fulfill more simple tasks such as reading individual sensors (e.g., $FB_3$, cf. Table 1). Hence, the lower maturity values can be explained by the applied programming strategies in Company A in the case of smaller modules, which can cause lower maturity values as reflected by metric (1.1+).

On the other hand, the results in Tab. 3 are particularly interesting as they do not confirm indicator (2), which states that FBs from lower levels are expected to have a high maturity. However, the results in Tab. 3 only refer to one company. Therefore, indicator (2) should not be regarded as disproved, as previous studies showed that there are companies that change FBs at higher levels much more extensively due to different degrees of standardization [53].

**Table 4:** Maturity evaluation with standard deviation σ of the last two years sorted by architectural level (changeset numbers in the individual levels do not sum up to 856, because a changeset partly comprises files from different architectural levels leading to overlaps)

|  | Maturity | σ | #Change Sets |
|---|---|---|---|
| *All changesets* | *94%* | *0.082* | *856* |
| **Stations** | 94% | 0.071 | 451 |
| **Sub-systems** | 94% | 0.077 | 420 |
| **General Functions** | 93% | 0.10 | 129 |

Summarizing, MICOSE4aPS reflects the investigated influencing factors on the way of implementing changes. However, usually, programming strategies are highly dependent on company-specific constraints. Therefore, the validity of R1.3 could only be confirmed for Company A.

### Validation of the metric's adaptability (R2)

The experts of Company A confirmed that MICOSE4aPS is scalable regarding the size of different FBs through the size factors $k_e$ and $k_l$ introduced in Section 4.1, and agreed that using the enlarged calculation rule distinguishing 69 possible change terms instead of the original eight terms [52], change criticalities can be differentiated on a fine-grained level. Hence, the adaptability of MICOSE4aPS regarding different FB sizes and change scopes and, therefore, the validity of R2.1 can be confirmed.

Recently, MICOSE4aPS has been adapted to apply to IEC 61131-3-LD in cooperation with a second Company B, i.e., a special-purpose machine manufacturer for the automotive, medical, and solar domain. To adapt the maturity calculation, a workshop with PLC programmers of Company B was conducted analogously to the case study introduced above. They agreed on the underlying assumptions and confirmed that the concept is transferable to IEC 61131-3-LD programs if adaptions are made, e.g., regarding operators' weighting or different PLC platforms. Additionally, the concept was presented to software engineers of two further companies from the field of packaging machinery, i.e., Company C, which mainly operates in the medical and pharmaceutical domain and received the second best results in a preceding case study [51], and Company D [57], which is a leading manufacturer for end-of-line-packaging machines. The approach was rated as reasonable in all companies and confirmed the great need for an objective maturity calculation for IEC 61131-3 software. In Company D, which, e.g., uses a different PLC programming environment compared to Company A, metric (1.1+) was implemented, which confirms the concept's adaptability. All in all, R2.2 is fulfilled.

**Table 5:** Overview of the Fulfilment of Requirements by MICOSE4aPS

| Requirement | Sub-requirement | Summary and findings | Evaluation results | Validity of results |
|---|---|---|---|---|
| MICOSE4aPS should deliver reliable results for | MICOSE4aPS reliably leads to high maturity values in case of changes with low criticality (R1.1). | Evaluation with module developers of Company A confirmed the reliable identification of non-critical changes by high maturity values |  |  |
| industrial control code (R1) | MICOSE4aPS reliably leads to low maturity values in case of changes with high criticality (R1.2). | Evaluation with module developers of Company A confirmed the reliable identification of critical changes by low maturity values | 🟢 | 🟢 |
| | MICOSE4aPS should be able to reflect different factors that influence change impact (R1.3). | MICOSE4aPS reflects influences on the change impact resulting from applied programming strategy, change type, number of changes made to the POU before, lifecycle phase, and architectural level. | 🟢 | 🟡 |
| MICOSE4aPS should be adaptable for different boundary conditions (R2) | MICOSE4aPS should be scalable to different sizes of POUs and scopes of change (R2.1). | Scalability of MICOSE4aPS by introducing scaling and weighting factors and by considering 69 possible operator changes. | 🟢 | 🟢 |
| | MICOSE4aPS should be adaptable to different company-specific constraints (R2.2) | General applicability in other companies and programming languages demonstrated for four companies. | 🟢 | 🟡 |
| The application of MICOSE4aPS in an industrial context should support the module developer in a beneficial way (R3) | The results of the maturity metric and their visualization should be understandable for module developers (R3.1) | PLC software developers of Company A confirmed understandability of MICOSE4aPS. | 🟢 | 🟢 |
| | MICOSE4aPS can be integrated into a company's software development workflow (R3.2) | Integrability of MICOSE4aPS confirmed for two companies. | 🟢 | 🟡 |

**Validation of the metric's industrial benefit (R3)**

The PLC software developers agreed that the visualization of maturity values (cf. Figure 9) is intuitively understandable and useful (R3.1). It is clear how previous changes are composed by mouse-over events and what the different change types mean. Therefore, the maturity calculation can be directly used by the PLC software developers to decide whether they should commit a change or whether they should reconsider the change implementation due to high criticality.

The approach is directly integrated into the company's workflow for software development as it is linked to the version management tool (R3.2) (cf. Section 5.1). The feedback dialogue calculation time depends on the time necessary to calculate the maturity of the committed changeset. The time required to open MongoDB and store the feedback value is not relevant due to the concurrent implementation. The calculation takes 4.5 seconds on average, making it acceptable for the PLC software developers not to be delayed in their regular work. The resulting metric was also implemented for Company D by executing the calculation during the nighttime and evaluating the maturity protocol the next morning. This proves that there are different ways to integrate MICOSE4aPS into the software development workflow of different companies and MICOSE4aPS fulfills R3.2.

## 5.3 Discussion of Results

Assessing a POU's reusability is a complex challenge dependent on a multitude of various influencing factors. MICOSE4aPS targets the maturity of control software library modules by quantifying the influence of a change on a high-quality POU's reusability, which has been tried and tested beforehand by module developers. This assumes that the POU already show low coupling and standardized interfaces, which are seen as prerequisites for a module to be included to a library. In summary, MICOSE4aPS has proven to be a reliable, beneficial approach for assessing and visualizing the maturity of control software library modules in an industrial environment.

However, some of the requirements were evaluated using only one company's software, leading to the fact that company-specific constraints influence the evaluation, e.g., the software architecture (cf. Figure 9) or programming strategies depending on an FB's size or reliability. Therefore, it is necessary to investigate whether MICOSE4aPS meets these requirements in other companies in future work.

MICOSE4aPS does not yet identify metadata such as the change type or the lifecycle phase. However, it must be extracted separately from the version management tool or asked by the developers to interpret the results. In future work, this could be addressed by additional entries to the exported JSON files and evaluate this information, e.g., using regular expressions to search for naming conventions.

MICOSE4aPS reflects the impact of different ways of implementing changes depending, e.g., on the FB size (cf. R1.3), leading to the challenge that a programmer who wants to receive a high maturity value for his change could "trick" the approach by splitting a large change into many small commits. On the other hand, a consultation with PLC programmers from Company A revealed that it is generally desirable to distribute large changes over several small commits to minimize the risk for errors. Therefore, a high maturity for this procedure is appropriate. Nevertheless, after completing a new feature, MICOSE4aPS is required to identify that the software needs to be extensively tested. Therefore, a correct interpretation is important by comparing the POU states before initiating and finishing a critical change.

In its current state, MICOSE4aPS focuses on assessing the changes within an individual POU without considering potential cross-effects on calling POUs. However, as even small changes in the implementation of a POU can strongly affect the POUs connected to it, it is planned also to include the call frequency and connections between POUs in future work, e.g., by considering information flow metrics such as Fan-In or Fan-Out [25] as an indicator for a POU's incoming and outgoing dependencies (cf. Table 2). On the other hand,

MICOSE4aPS currently does not yet cover concepts of the object-oriented extension of the IEC 61131-3 defining, e.g., methods to encapsulate different functionalities of an FB. Because of the rising importance of the object-oriented programming paradigm in PLC programming, we aim to consider these concepts in MICOSE4aPS and analyze the effect of changing a method's implementation on an FB's reusability in future work.

MICOSE4aPS can be adapted to further companies, e.g., by specifying the company-specific parameter $s_2$ (cf. equation (6)). In case another programming language than IEC 61131-3-ST is used, the operators of the applied language must be weighted analogously based on expert feedback as demonstrated for Company B. However, as MICOSE4aPS can only be applied to companies that document the change history of POUs. In case a company uses a tool other than Mercurial, the implementation of MICOSE4aPS requires adaptions of the interface to the applied system (cf. R3.2).

Overall, MICOSE4aPS provides valuable support in industrial control software development and yields potential for enlargements in future work.

# 6 Conclusion and future work

Although there is a variety of different approaches available to optimize reusability and maintainability of software in different disciplines, there is still a gap to bridge for control software in aPS: The special characteristics such as hard real-time requirements, different qualification levels of the application developers, or short-term changes during commissioning cause that established approaches from computer science are not transferable without adjustments. Therefore, the paper introduces the new approach MICOSE4aPS for reliably calculating the maturity of control software library modules in machine and plant automation as an objective indicator for the reusability of these modules without risk. Systematic reuse of mature modules can be enhanced, which is a significant challenge in industrial companies from the field of aPS. An industrial case study could confirm the applicability and benefits of MICOSE4aPS: The calculation rule is scalable for large module sizes, and the concept can be directly integrated into the software development workflow, e.g., by linking it to a version management tool. Experts of four companies confirmed the benefit of a reliable, objective measure to determine and visualize the maturity of POUs in addition to their gut feeling. MICOSE4aPS can identify FBs affected frequently by comprehensive changes and, hence, require specified testing procedures. MICOSE4aPS was used to analyze correlations between maturity and change number and type, architectural level, lifecycle phase, and programming strategies. Experts proved the company-specific adaptability of MICOSE4aPS in four companies from machine and plant manufacturing, who confirmed the approach's benefits.

The evaluation of the concept revealed the high potential of MICOSE4aPS. To ensure full applicability for all five languages of the IEC 61131-3, further adaptions of the concept are required and addressed in future research. Moreover, it is planned to introduce further terms to refine the calculation rule, e.g., successful tests of changes, the instantiation frequency of POUs, or influences on calling POUs.

## Acknowledgment

The Institute of Automation and Information Systems thanks MULTIVAC Sepp Haggenmüller SE & Co. KG for the great support in the evaluation of the derived maturity metrics by providing code examples and tool support and by enriching the analysis with valuable expert feedback.